\documentclass{ws-ijmpb}

\newtheorem{conjecture}[theorem]{Conjecture}

\def\dps{\displaystyle}

\begin{document}

\markboth{M. T. Batchelor, J. de Gier and B. Nienhuis}
{The Rotor Model and Combinatorics}

\catchline{}{}{}

\title{THE ROTOR MODEL AND COMBINATORICS}

\author{M. T. BATCHELOR and J. de GIER} 

\address{Centre for Mathematics and its Applications,\\and\\ 
Department of Theoretical Physics, I.A.S.,\\
Australian National University, Canberra ACT 0200, Australia}

\author{B. NIENHUIS}

\address{Instituut voor Theoretische Fysica, Universiteit van
Amsterdam,\\  1018XE Amsterdam, The Netherlands}

\maketitle

\pub{Received (received date)}{Revised (revised date)}

\begin{abstract}
We examine the groundstate wavefunction of the
rotor model for different boundary conditions.
Three conjectures are made on the appearance of numbers enumerating 
alternating sign matrices.
In addition to those occurring in the O($n=1$) model we find 
the number $A_{\rm V}(2m+1;3)$, which 3-enumerates vertically
symmetric alternating sign matrices.
\end{abstract}

\section{Introduction}	

The XXZ Heisenberg spin chain and the related six-vertex model
stand as central pillars in the study of exactly solved 
models in statistical mechanics.\cite{1,2}
It has been known for many years that, with appropriate boundary
conditions, their groundstate energy is trivial at the particular
anisotropy value $\Delta = -1/2$. 
Only recently has it been realised that the corresponding groundstate
wavefunction possesses some rather remarkable properties.\cite{3,4,5} 
These observations extend to the related O($n$) loop model\,\cite{6,7} at
$n=1$.\cite{4,8,9,10}

Consider first the periodic antiferromagnetic XXZ chain 
\begin{equation}
H = - \frac{1}{2} \sum_{j=1}^{N} \left( \sigma_j^x \sigma_{j+1}^x +
\sigma_j^y \sigma_{j+1}^y + \Delta \sigma_j^z \sigma_{j+1}^z \right),
\label{eq:XXZham}
\end{equation}
defined on an odd number $N$ of sites.
Here $(\sigma_j^x,\sigma_j^y,\sigma_j^z)$ are the Pauli spin
matrices acting at site $j$.
Normalize the smallest component of the groundstate wavefunction
to be unity.
Then at $\Delta = -1/2$ the largest component is 
conjectured to be given by\,\cite{3} 
\begin{equation}
A(m) = 
%
\prod_{j=0}^{m-1} \frac{(3j+1)!} {(m+j)!},
\label{eq:asm}
\end{equation}
for size $N=2m+1$.
The remarkable point being that $A(m)$ is
the number of $m \times m$ alternating sign matrices.\cite{11}
The resulting sequence $A(m) = 1, 2, 7, 42, 429, 7436 \ldots$ is also
known to count other combinatorial objects.\cite{12,13}
Moreover, these numbers appear in the sum of all the 
groundstate wavefunction components.
These observations remain to be proved.

An even number of sites and other boundary conditions 
have also been considered, both for the XXZ chain (twisted
and closed \footnote{The standard nomenclature for these bc's is open
bc, but since these bc's are spin-conserving in the XXZ chain or loop
reflecting in the O($n=1$) model we find the term closed bc more
appropriate, here reserving open bc for non-conserving boundary conditions.}
quantum symmetric bc's) and the O($n=1$) loop model (periodic and
closed bc's). 
These see the appearance of other well known numbers counting
alternating sign matrices and related objects in different symmetry
classes.
For example, with the smallest component of the
groundstate wavefunction again unity, the O($n=1$) loop model with
closed boundary conditions has largest component given by
$A_{\rm V}(2m-1)$ for $N=2m-1$ and $N_8(2m)$ for $N=2m$.
Here 
\begin{equation}
A_{\rm V}(2m+1) = \prod_{j=0}^{m-1} (3j+2) \frac{(2j+1)!(6j+3)!}
{(4j+2)!(4j+3)!} 
\end{equation}
is the number of $(2m+1) \times (2m+1)$
vertically symmetric alternating sign matrices and
\begin{equation}
N_8(2m) = \prod_{j=0}^{m-1} (3j+1) \frac{(2j)!(6j)!}
{(4j)!(4j+1)!}
\end{equation}
is the number of cyclically symmetric transpose complement
plane partitions.
The number $N_8(2m)$ is conjectured to be $A_{\rm VH}(4m+1)/A_{\rm
V}(2m+1)$, where $A_{\rm VH}(4m+1)$ is the number of $(4m+1) \times
(4m+1)$ vertically and horizontally symmetric alternating sign
matrices\,\cite{14,15}. Another quantity, which appears for periodic
boundary conditions, is 
\begin{equation}
A_{\rm HT}(2m) = A(m)^2 \prod_{j=0}^{m-1} \frac{3j+2}{3j+1},
\end{equation}
the number of $2m \times 2m$ half turn symmetric alternating sign
matrices.

Further developments include the combinatorial
interpretation of the elements of the O($n=1$) loop
model wavefunction in terms of link patterns\,\cite{8,9,10} and the 
relation to a one-dimensional stochastic process\,\cite{10}.
There has been some progress attempting to prove these conjectures
using Bethe Ansatz techniques.\cite{16,17}  

In this paper, we examine the groundstate wavefunction of the
rotor model\,\cite{18} discussed by Martins and Nienhuis.
The rotor model is based on a variant of the Temperley-Lieb algebra,
which underpins the six-vertex model, the O($n$) model and the 
critical $Q$-state Potts model.\cite{1,2,19}
The rotor model is defined in Section 2, with our results presented
in Section 3.

Here we see the appearance of another number, $A_{\rm V}(2m+1;3)$, 
which is the 3-enumeration of $(2m+1)\times
(2m+1)$ vertically symmetric alternating sign matrices, or
equivalently, the number of vertically symmetric 6-vertex
configurations with domain wall boundary conditions and $\Delta = -1/2$.
It is given by
\begin{equation}
A_{\rm V}(2m+1;3) = \frac{3^{m(m-3)/2}}{2^m}
\prod_{j=1}^m \frac{(j-1)!(3j)!}{j (2j-1)!^2} = 1, 5, 126,
16038, \ldots.
\end{equation}
In general, the $x$-enumeration of alternating sign matrices in the
terminology of Kuperberg\,\cite{15}, is equivalent to the enumeration of
six-vertex configurations with domain wall boundaries with $\Delta=
1-x/2$ and at the symmetric point with respect to the spectral
parameter.

We give some concluding remarks in Section 4.

\section{The rotor model}
\begin{figure}[h]
\centerline{
\begin{picture}(220,60)
\put(0,20){\epsfxsize=220pt\epsfbox{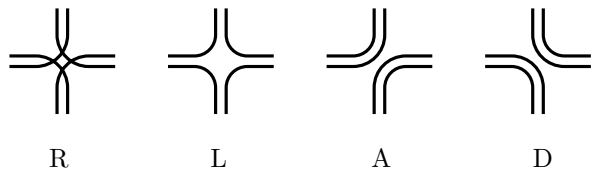}}
\put(15,0){R}
\put(76,0){L}
\put(137,0){A}
\put(198,0){D}
\end{picture}}
\caption{Vertices of the rotor model.}
\label{vertices}
\end{figure}

We suggest that the remarkable observations of this O($n=1$) model are
related to the combination of two key properties, namely
solvability and the absence of finite size corrections to the
groundstate energy. Now the O($n=1$) model is not unique in this
combination. Recently Martins and Nienhuis\,\cite{18} introduced a
model that shares the same two properties. In this so-called rotor
model a set of loops covers all the edges of the square lattice
precisely twice. At the vertices all the loops make a turn of $\pi/2$
which permits four types of vertices as displayed in Figure \ref{vertices}.

A natural interpretation is that the loops are trajectories of
particles, and that the two loop segments visiting the same edge are
traversed in opposite directions. Thus the four kinds of vertices
shown in Figure \ref{vertices} behave as scatterers: right (R) and
left (L) rotors, at which the particles always turn right and left
respectively, and ascending (A) and descending (D) diagonal mirrors at
which the particles get reflected.
To clearly display the scatterers we propose that the
particles always follow the left hand side of the road, as is customary
in Australia where this paper was conceived. 

In a different interpretation the two loop segments at the same edge 
are the trajectories of different kinds of particles, traversed in
either direction.
Then the scatterers can all be interpreted as double mirrors on each
site, each reflecting one kind of particle and transmitting the
other.
At the R and L sites these mirrors are placed crosswise, AD
and DA respectively, while at the original ascending and descending
mirrors, the double mirrors are placed parallel, AA and DD respectively.
This alternate interpretation will not affect the distributions of
trajectories in an infinite system, but it will result in changes on
some finite systems.

\begin{figure}[h]
\centerline{
\begin{picture}(127,49)
\put(0,20){\epsfxsize=127pt\epsfbox{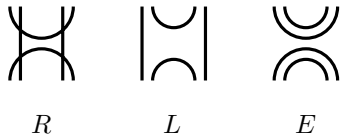}}
\put(10,0){$R$}
\put(60,0){$L$}
\put(110,0){$E$}
\end{picture}}
\caption{Generators.}
\label{generators}
\end{figure}

Martins and Nienhuis solved this model by means of the Yang-Baxter
equation when these scatterers occur with the respective weights
\begin{equation}
\begin{array}{l}
\dps \omega_{\rm R} = \omega_{\rm L} = \sin u \cos(2\pi/3-u), \\
\dps \omega_{\rm A} = \sin(\pi/3-u)\cos(2\pi/3-u), \\
\dps \omega_{\rm D} = -\sin u \cos(\pi/3-u). \\
\end{array}
\end{equation}
independently at each vertex.
In this paper we consider this model with periodic boundary conditions 
(pbc) and with closed boundaries at which the trajectories are reflected.
We will be interested in the structure of the groundstate
eigenvector. Since the transfermatrix as a function of $u$ forms a
commuting family, the groundstate is independent of $u$. Then it is
convenient to consider the Hamiltonian, found (up to a constant) as
the logarithmic derivative of the transfer matrix with respect to $u$
at $u=0$: 
\begin{equation}
H = \sum_{i} 3 - R_i - L_i - E_i. \label{eq:rotorham}
\end{equation}
For system size $N$ the operators $R$, $L$ and $E$ are shown in terms
of the loops in Figure \ref{generators}. 

Martins and Nienhuis showed that the operators $L_{2i}$ and $R_{2i-1}$
generate a Temperley-Lieb (TL) algebra, and so do the operators
$L_{2i-1}$ and $R_{2i}$. In periodic systems of even size, and in
bounded systems these two TL algebras commute with each other. 
What changes the physics is the presence in the Hamiltonian of the
term $E_i = R_i L_i$. Also the $E_i$ by themselves generate a TL algebra.
In odd, periodic systems the odd and even sites cannot be
distinguished. In this case the $L$ and the $R$ together form a TL
algebra of $2 N$ sites. 

When the system is odd and periodic, the interpretation of the $R$ and
$L$ vertices as rotors or alternatively as crossing mirrors, will
naturally result in different pbc. The
rotor interpretation permits closed trajectories that wind the
cylinder twice. In the alternative interpretation no closed winding
trajectories are possible, and the odd system must have two unmatched
terminals. In this paper we follow the latter interpretation. 

The states of the model are the pairings of those terminals that are
connected by a trajectory in the `past' half of the strip or cylinder.
When the system is periodic, one may distinguish the side of the
cylinder along which the trajectory runs: a connection between site 1
and site $N$ may pass all sites ${2,\ldots N-1}$, or it may simply 
connect site $N$ to site $N+1$ which is identified to 1. These two
states can be distinguished, in which case we speak of pbc
{\it per se}, or they may be identified, for which
we reserve the phrase pbc with identified connectivities.
\section{Results for the groundstate wavefunction}
The groundstate wavefunction of the Hamiltonian (\ref{eq:rotorham})
satisfies the eigenvalue equation
$
H \psi_0 = 0.
$
In this section we formulate three conjectures regarding $\psi_0$ for
the different types of boundary conditions discussed in Section 2. 
\begin{conjecture} For closed boundary conditions, if the smallest
element of the rotor model groundstate wavefunction for $N=2m-1$ is
normalized to $A_{\rm V}(2m-1;3)$, then all of its elements are
integers and the sum of its elements is given by $S(2m-1) =
3^{(m-1)^2} N_8(2m)$. For $N=2m$, normalize the groundstate
wavefunction to the smallest integer such that all elements are
integers, the sum of the elements is given by $S(2m) = 3^{2\theta_m}
A_{\rm V}(2m+1)$, where $\theta_m = 0,1,3,6,9 = \lfloor (m-1)(m+2)/3
\rfloor$ for $m=1,\ldots,5$. 
\end{conjecture}
 
This conjecture is based on the results presented in 
Table 1 and was checked up to $N=10$.

\begin{conjecture} For periodic boundary conditions, normalize the
smallest element of the rotor model groundstate wavefunction to the
smallest integer such that all elements are integer. The sum of its
elements is then given by $S(2m-1) = 3^{3m} A_{\rm V}(2m+1;3)^2$ 
for odd system sizes and by
$S(2m) = 3^{m^2} A_{\rm HT}(2m)$
for even system sizes. 
\end{conjecture}
 
This conjecture is based on the results presented in 
Table 2 and was checked up to $N=9$.
 
\begin{conjecture} For periodic boundary conditions and identified
connectivities, normalize the smallest element of the rotor model
groundstate wavefunction to the smallest integer such that all
elements are integer. The sum of its elements is then given by $S(2m)
= 3^{\theta_m} A(m)$, where $\theta_m = 0,1,3,6,9,13 = \lfloor
(m-1)(m+2)/3 \rfloor$ for $m=1,\ldots,6$. 
\end{conjecture}

This conjecture is based on the results presented in
Table 3 and was checked up to $N=12$.

\begin{table}[htbp]
\ttbl{30pc}{Groundstate wavefunctions of the rotor model with closed
boundaries. Note that by $\psi_0 = (2,1)$ with multiplicity $(2,2)$ we
mean $\psi_0 = (2,2,1,1)$.}
{\begin{tabular}{ccp{5cm}p{3cm}l}\\
\hline
$N$  & $m$ & $\psi_0$  & {\rm multiplicity} & $S_{N}^{(1)}$    \\ \hline
  1  &1    & (1) & (1)   & 1    \\
  2  &1    & (1) & (1)   & 1    \\
  3  &2    & (2,1) & (2,2)   & 6    \\
  4  &2    & (14,5,4) & (1,1,2)   & 27    \\
  5  &3    & (113, 111, 55, 31, 25, 21, 19, 11, 5)
           & (2, 1, 4, 2, 4, 2, 4, 4, 2)   & 891    \\
  6  &3    & (4760, 1440, 1192, 1028, 601, 565, 326, 310, 126, 121,
              86) & (1, 2, 4, 1, 4, 2, 2, 2, 1, 2, 4)   & 18954    \\
\hline
\end{tabular}}
\end{table}

\begin{table}[htbp]
\ttbl{30pc}{Groundstate wavefunctions of the rotor model with periodic
boundaries.}
{\begin{tabular}{ccp{5cm}p{3cm}l}\\
\hline
$N$  & $m$ & $\psi_0$  & {\rm multiplicity} & $S_{N}^{(1)}$    \\ \hline
  1  &1    & (1) & (1)   & 1    \\
  2  &1    & (2,1) & (2,2)   & 6    \\
  3  &2    & (5,2) & (3,6)   & 27    \\
  4  &2    & (118, 35, 25, 22, 20, 5, 4)
           & (2, 2, 8, 4, 8, 8, 4)   & 810    \\
  5  &3    & (1036, 463, 208, 143, 127, 122, 65, 22, 10)
           & (5, 10, 10, 20, 5, 10, 20, 10, 10)   & 18225    \\
\hline
\end{tabular}}
\end{table}

\begin{table}[htbp]
\ttbl{30pc}{Groundstate wavefunctions of the rotor model with periodic 
boundaries and identified connectivities.}
{\begin{tabular}{ccp{5cm}p{3cm}l}\\
\hline
$N$  & $m$ & $\psi_0$  & {\rm multiplicity} & $S_{N}^{(1)}$    \\ \hline
  2  &1    & (1) & (1)   & 1    \\
  4  &2    & (2,1) & (2,2)   & 6    \\
  6  &3    & (26, 9, 7, 2) & (2, 3, 14, 6)   & 189    \\
  8  &4    & (1798, 486, 410, 267, 234, 232, 165, 106, 90, 81, 76, 70,
              56, 45, 20, 9, 4)
           & (2, 8, 16, 2, 16, 16, 8, 16, 4, 16, 8, 8, 16, 32,
              16, 8, 4)   &  30618   \\
\hline
\end{tabular}}
\end{table}

\section{Discussion}

In this paper we have examined the groundstate wavefunction of the
rotor model for three different boundary conditions.
As for the O($n=1$) model, 
numbers known to enumerate equally weighted alternating sign matrices
appear in the normalization of the wavefunction. 
For the rotor model we also see the number $A_{\rm V}(2m+1;3)$,
enumerating alternating sign matrices in which the minus signs
have weight 3.\cite{15}

We find it quite surprising that the conjectures in Section 3
can be formulated at all.
They are a result of the normalizations factoring into 
relatively small primes and thus enabling their recognition.  
This property appears to be absent for other boundary conditions,
for example, pbc in the
rotor interpretation for odd system sizes.
It is even more remarkable that these numbers have 
a well known combinatorial meaning.

\section*{Acknowledgements}

It is a great pleasure to dedicate this paper to Fred Wu on the
occasion of his 70th birthday.
This work has been supported by the Australian Research Council and 
the Dutch foundation `Fundamenteel Onderzoek der Materie' (FOM).


\clearpage
\begin{thebibliography}{0}

\bibitem{1}
E. H. Lieb and F. Y. Wu, in {\it Phase Transitions and
Critical Phenomena, Vol 1}, eds. C. Domb and M. S. Green 
(Academic Press, London, 1972) p. 331.

\bibitem{2}
R. J. Baxter, {\it Exactly Solved Models in Statistical Mechanics} 
(Academic Press, London, 1982).

\bibitem{3} 
A. V. Razumov and Yu. G. Stroganov, 
{\it J. Phys. A} {\bf 34}, 3185 (2001).

\bibitem{4} 
M. T.~Batchelor, J.~de Gier and B.~Nienhuis,
{\it J. Phys. A} {\bf 34}, L265 (2001).
 
\bibitem{5} A. V. Razumov and Yu. G.~Stroganov,
{\it J. Phys. A} {\bf 34}, 5335 (2001).

\bibitem{6}
R. J. Baxter, S. B. Kelland and F. Y. Wu, 
{\it J. Phys. A} {\bf 9}, 397 (1976). 

\bibitem{7} H. W. J.~Bl\"ote and B.~Nienhuis,
{\it J. Phys. A} {\bf 22}, 1415 (1989).

\bibitem{8} 
A. V.~Razumov and Yu. G.~Stroganov,
``Combinatorial nature of ground state vector of O(1) loop model'',
arXiv:math.CO/0104216.
 
\bibitem{9} 
A. V.~Razumov and Yu. G.~Stroganov,
``O(1) loop model with different boundary conditions and
symmetry classes of alternating-sign matrices'',
arXiv:cond-mat/0108103.

\bibitem{10} 
P. A.~Pearce, V.~Rittenberg and J.~de~Gier,
``Critical Q=1 Potts model and Temperley-Lieb stochastic processes'',
arXiv:cond-mat/0108051.

\bibitem{11} W.H.~Mills, D.P.~Robbins and H.~Rumsey,
{\it J. Combin. Theory A} {\bf 34}, 340 (1983).

\bibitem{12} D. M.~Bressoud, {\it Proofs and Confirmations -- The
story of the alternating sign matrix conjecture}
(Cambridge University Press, 1999).

\bibitem{13}
J.~Propp, ``The many faces of alternating sign matrices'', preprint.

\bibitem{14}
D. P. Robbins, ``Symmetry classes of alternating sign matrices'',\\
arXiv:math.CO/0008045.

\bibitem{15} 
G.~Kuperberg, ``Symmetry classes of alternating-sign
matrices under one roof'', arXiv:math.CO/0008184.

\bibitem{16}
J. de Gier, M. T. Batchelor, B. Nienhuis and S. Mitra,
``The XXZ spin chain at $\Delta=- \frac{1}{2}$:
Elementary symmetric functions, Schur functions and determinants'',
arXiv:math-ph/0110011.

\bibitem{17}
N. Kitanine, J. M. Maillet, N. A. Slavnov and V. Terras, ``Emptiness
formation probability of the XXZ spin-$\frac12$ Heisenberg chain at
$\Delta=\frac12$'', arXiv:hep-th/0201134.
 
\bibitem{18}
M. J. Martins and B. Nienhuis, 
{\it J. Phys. A} {\bf 31}, L723 (1998).

\bibitem{19}
F. Y. Wu, {\it Rev. Mod. Phys.} {\bf 54}, 235 (1982). 

\end{thebibliography}
\end{document}